\documentstyle[12pt]{article}

\begin{document}

\begin{titlepage}
\vspace{1cm}

\begin{centering}

{\Large \bf Fermions and  noncommutative theories  }

\vspace{.5cm}
\vspace{1cm}

{\large Ricardo Amorim }

\vspace{0.5cm}

 Instituto de F\'{\i}sica, Universidade Federal
do Rio de Janeiro,\\
Caixa Postal 68528, 21945-970  Rio de Janeiro, Brazil\\[1.5ex]
\vspace{1cm}

\begin{abstract}
By using a framework where the object of noncommutativity $\theta^{\mu\nu}$ represents independent degrees of freedom, we study the symmetry properties of
an extended $x+\theta$ space-time, given by the group $P$', which has the Poincar\'{e} group $P$ as a subgroup. In this process we use the minimal canonical extension of the Doplicher, Fredenhagen and Roberts algebra. It is also proposed a generalized Dirac equation, where the fermionic field depends not only  on the ordinary coordinates but  on $\theta^{\mu\nu}$ as well.  The dynamical symmetry content of such  fermionic theory is discussed, and we show that its action is invariant under  $\cal P$'. 
\end{abstract}

\end{centering}

\vspace{1cm}


\vfill

\noindent{\tt amorim@if.ufrj.br}
\end{titlepage}

\pagebreak

\renewcommand{\theequation}{\arabic{equation}}
\setcounter{equation}{0}

\bigskip 
Since the seminal work of Snyder   \cite{Snyder}, space-time noncommutativity has deserved a great amount of study. Nowadays the subject is mainly associated with strings    \cite{Strings}, noncommutative field theories(NCFT's)   \cite{NCFT} and gravity   \cite{QG}, which can be related 
  \cite{Hull,SW}.  In the fundamental noncommutativity relation

\begin{equation}
\label{02}
[{\mathbf x}^\mu,{\mathbf x}^\nu] = i {\mathbf \theta}^{\mu\nu}
\end{equation} 

\noindent the object of noncommutativity  ${\mathbf \theta}^{\mu\nu}$ is usually considered as a constant matrix, what breaks Lorentz symmetry   \cite{NCFT}. Actually, if strings have their end points on D-branes, in the presence of  a constant antisymmetric tensor field background, this kind of canonical noncommutativity effectively arises. Recently it was discovered that
the structure (\ref{02}) is compatible with the twisted Poincar\'{e} symmetry \cite{twisted}, which can be an important ingredient in the construction of NCFT's. In a different perspective,
 ${\mathbf \theta}^{\mu\nu}$  can also be considered as an independent operator   \cite{Carlson}-  \cite{Saxell}, which permits to construct true Lorentz invariant theories. As too little is known about Physics at Planck scale, it seems interesting to study different structures which could arise in a very high energy regime.
The last works cited above are based on a contraction of the  Snyder's algebra  or directly on the  
Doplicher, Fredenhagen and Roberts (DFR) algebra   \cite{DFR}. The DFR algebra
 assumes, besides (\ref{02}), the structure

\begin{eqnarray}
\label{03}
&&[{\mathbf x}^\mu,{\mathbf \theta}^{\alpha\beta}] =0\nonumber\\
&&[{\mathbf \theta}^{\mu\nu},{\mathbf \theta}^{\alpha\beta}]=0
\end{eqnarray}

\noindent although more general settings could be assumed \cite{Doplicher}. The authors of   \cite{DFR} also propose the quantum conditions

\begin{eqnarray}
\label{04}
&&{\mathbf \theta}_{\mu\nu}{\mathbf \theta}^{\mu\nu} =0\nonumber\\
&&({1\over4}\star{\mathbf \theta}^{\mu\nu}{\mathbf \theta}_{\mu\nu})^2=\lambda_P^8
\end{eqnarray}

\noindent where $\star{\mathbf \theta}_{\mu\nu}={1\over2}\epsilon_{\mu\nu\rho\sigma}{\mathbf \theta}^{\rho\sigma}$and $\lambda_P$ is the Planck length.
Their theory illuminates localizability in a quantum space-time, which has to be extended in order to include the object of noncommutativity as an independent set of coordinates. The same occurs in Refs.    \cite{Carlson}-  \cite{Saxell}.

In two recent works   \cite{Amorim1,Amorim4}, the author proposes a minimal canonical extension of the DFR algebra in order to implement, in a noncommutative
quantum mechanics (NCQM) framework   \cite{NCQM}, Poincar\'{e} invariance as a dynamical symmetry   \cite{Iorio}. Of course this represents one among several possibilities of incorporating noncommutativity in quantum theories. In \cite{Amorim4} not only
the coordinates ${\mathbf x}^\mu$ and their conjugate momenta ${\mathbf p}_\mu$ are operators acting in a Hilbert space ${\cal H}$, but also  
$\theta^{\mu\nu}$ and their canonical momenta ${\mathbf \pi}_{\mu\nu}$ are considered as Hilbert space operators as well. Besides (\ref{02}), (\ref{03}), the proposed extension of the DFR algebra is given by

\begin{eqnarray}
\label{i2}
&&[{\mathbf x}^\mu,{\mathbf p}_\nu ] = i\delta^{\mu}_{\nu}\nonumber\\
&&[{\mathbf p}_\mu,{\mathbf p}_\nu ] = 0\nonumber\\
&&[{\mathbf \theta}^{\mu\nu},{\mathbf \pi}_{\rho\sigma}]=i\delta^{\mu\nu}_{\,\,\,\rho\sigma} \nonumber\\
&&[{\mathbf \pi}_{\mu\nu},{\mathbf \pi}_{\rho\sigma}]=0\nonumber\\
&&[{\mathbf p}_\mu,{\mathbf \theta}^{\rho\sigma}]=0\nonumber\\
&&[{\mathbf p}_\mu,{\mathbf \pi}_{\rho\sigma}]=0\nonumber\\
&&[{\mathbf x}^\mu,{\mathbf \pi}_{\rho\sigma}]=-{i\over2}\delta^{\mu\nu}_{\,\,\,\rho\sigma}p_\nu
\end{eqnarray}

\noindent where $\delta^{\mu\nu}_{\,\,\,\,\rho\sigma}=\delta^\mu_\rho\delta^\nu_\sigma-\delta^\mu_\sigma\delta^\nu_\rho$. The relations above are consistent under all possible Jacobi identities. 

\bigskip

Now it is possible to adopt   \cite{Gracia}

\begin{equation}
\label{i16}
{ \mathbf M}^{\mu\nu}= { \mathbf X}^\mu{\mathbf p}^\nu-{\mathbf X}^\nu{\mathbf p}^\mu-{\mathbf \theta}^{\mu\sigma}{\mathbf \pi}_\sigma^{\,\,\nu}+{\mathbf \theta}^{\nu\sigma}{\mathbf \pi}_\sigma^{\,\,\mu}
\end{equation}

\bigskip\noindent where   \cite{NCQM}

\begin{equation}
\label{i5}
{\mathbf X}^\mu={\mathbf x}^\mu+{1\over2}{\mathbf\theta}^{\mu\nu}{\mathbf p}_\nu
\end{equation}

\noindent  as the  generator of the Lorentz group. It  closes in the appropriate algebra

\begin{equation}
\label{i17}
[{\mathbf M}^{\mu\nu},{\mathbf M}^{\rho\sigma}]=i\eta^{\mu\sigma}{\mathbf M}^{\rho\nu}-i\eta^{\nu\sigma}{\mathbf M}^{\rho\mu}-i\eta^{\mu\rho}{\mathbf M}^{\sigma\nu}+i\eta^{\nu\rho}{\mathbf M}^{\sigma\mu}
\end{equation}
 
\bigskip\noindent and generates the  expected Lorentz transformations on the Hilbert space operators. Observe that the usual  form of the ordinary Lorentz generator, given by ${ \mathbf l}^{\mu\nu}= { \mathbf x}^\mu{\mathbf p}^\nu-{\mathbf x}^\nu{\mathbf p}^\mu$, fails to close in an algebra if (\ref{02}) is adopted.

Now, let us define the dynamical transformation of an arbitrary operator ${\mathbf A}$ in $\cal H$ by
$\delta {\mathbf A}=i[ {\mathbf A}, {\mathbf G}]$. 
In the above expression let us choose
$ {\mathbf G}={1\over2}\omega_{\mu\nu}{\mathbf M}^{\mu\nu}-a^\mu{\mathbf p}_\mu+{1\over2}b^{\mu\nu}{\mathbf \pi}_{\mu\nu}$, where $\omega^{\mu\nu}=-\omega^{\nu\mu}$, $a^\mu$ and $b^{\mu\nu}=-b^{\nu\mu}$ are infinitesimal parameters. We get

\begin{eqnarray}
\label{i19}
\delta {\mathbf x}^\mu&=&\omega ^\mu_{\,\,\,\,\nu}{\mathbf x}^\nu+a^\mu+{1\over2}b^{\mu\nu}p_\nu\nonumber\\ 
\delta {\mathbf X}^\mu&=&\omega ^\mu_{\,\,\,\,\nu}{\mathbf X}^\nu+a^\mu\nonumber\\ 
\delta{\mathbf p}_\mu&=&\omega _\mu^{\,\,\,\,\nu}{\mathbf p}_\nu\nonumber\\
\delta{\mathbf \theta}^{\mu\nu}&=&\omega ^\mu_{\,\,\,\,\rho}{\mathbf \theta}^{\rho\nu}+ \omega ^\nu_{\,\,\,\,\rho}{\mathbf \theta}^{\mu\rho}+b^{\mu\nu}\\
\delta{\mathbf \pi}_{\mu\nu}&=&\omega _\mu^{\,\,\,\,\rho}{\mathbf \pi}_{\rho\nu}+ \omega _\nu^{\,\,\,\,\rho}{\mathbf \pi}_{\mu\rho}\nonumber\\
\delta {\mathbf M}^{\mu\nu}&=&\omega ^\mu_{\,\,\,\,\rho}{\mathbf M}^{\rho\nu}+ \omega ^\nu_{\,\,\,\,\rho}{\mathbf M}^{\mu\rho}+a^\mu{\mathbf p}^\nu-a^\nu{\mathbf p}^\mu+b^{\mu\rho}{\mathbf \pi}_\rho^{\,\,\,\,\nu}+ b^{\nu\rho}{\mathbf \pi}_{\,\,\,\rho}^{\mu}\nonumber
\end{eqnarray}

\noindent which generalizes the action of the Poincar\'{e} group $P$ in order to include $\theta$ translations. Let us refer to this group as $P$'. The $P$' transformations  close in an algebra. As can be verified, 

\begin{equation}\label{xxx}
[\delta_2,\delta_1]\,{\mathbf y}=\delta_3\,{\mathbf y}
\end{equation}

\noindent where the parameters composition rule is given by

\begin{eqnarray}
\label{i19b}
&&\omega^\mu_{3\,\,\,\,\nu}=\omega^\mu_{1\,\,\,\,\alpha}\omega^\alpha_{2\,\,\,\,\nu}-\omega^\mu_{2\,\,\,\,\alpha}\omega^\alpha_{1\,\,\,\,\nu}\nonumber\\
&&a_3^\mu=\omega^\mu_{1\,\,\,\nu}a_2^\nu-\omega^\mu_{2\,\,\,\nu}a_1^\nu\nonumber\\
&&b_3^{\mu\nu}=\omega^\mu_{1\,\,\,\rho}b_2^{\rho\nu}-\omega^\mu_{2\,\,\,\rho}b_1^{\rho\nu}-\omega^\nu_{1\,\,\,\rho}b_2^{\rho\mu}+
\omega^\nu_{2\,\,\,\rho}b_1^{\rho\mu}
\end{eqnarray}

To understand the symmetry content displayed in (\ref{i19}), we observe that the Hilbert space $\cal H$ can be written as the direct product of two disjoint subspaces, ${\cal H}={\cal H}_1\otimes\,{\cal H}_2$. The operators ${\mathbf X}^\mu$, ${\mathbf p}_\mu$ and 
${ \mathbf M}_1^{\mu\nu}= { \mathbf X}^\mu{\mathbf p}^\nu-{\mathbf X}^\nu{\mathbf p}^\mu$ act on ${\cal H}_1$, and ${\mathbf \theta}^{\mu\nu}$, ${\mathbf \pi}_{\mu\nu}$ and 
${ \mathbf M}_2^{\mu\nu}= -{\mathbf \theta}^{\mu\sigma}{\mathbf \pi}_\sigma^{\,\,\nu}+{\mathbf \theta}^{\nu\sigma}{\mathbf \pi}_\sigma^{\,\,\mu}$ act on ${\cal H}_2$. Both ${ \mathbf M}_1^{\mu\nu}$ and ${ \mathbf M}_2^{\mu\nu}$ satisfy the Lorentz algebra (\ref{i16}) and their symmetry properties can be read from (\ref{i19}). The unexpected transformation of ${\mathbf x}^\mu$ can be understood from (\ref{i5}), since ${\mathbf x}^\mu$ can be seen as a nonlinear combination of operators acting on ${\cal H}_1$ and ${\cal H}_2$. Now ${\mathbf p}_\mu$ and 
${ \mathbf M}_1^{\mu\nu}$ generate the usual Poincar\'{e} group $P$, which is the semidirect product of the four dimensional Lorentz grope $L$ and the  four dimensional translation group $T_4$. The transformation group $G$ acting on ${\cal H}_2$ can be seen as the semidirect product of the four dimensional Lorentz group and the six dimensional translation group $T_6$. As it is well known, $P$ has ${\mathbf C}_1={\mathbf p}^2$ and  ${\mathbf C}_2={\mathbf s}^2$ as invariant Casimir operators, where 
${\mathbf s}_\mu={1\over2}\epsilon_{\mu\nu\rho\sigma}{\mathbf M}_1^{\nu\rho}{\mathbf p}^\sigma$  is the Pauli-Lubanski vector.  
${\mathbf C}_3={\mathbf \pi}^2$ and ${\mathbf C}_4={ \mathbf M}_2^{\mu\nu}{\mathbf \pi}_{\mu\nu}$ are the Casimir operators of $G$.
A possible representation for $P$' can then be given by the $11\times11$ matrix

\begin{equation}\label{P'}
D_{P'}(\Lambda,A, B)=
\pmatrix{\Lambda^\mu_{\,\,\,\nu}&0&A^\mu\cr
0&\Lambda^{[\mu}_{\,\,\,\alpha}\Lambda^{\nu]}_\beta&B^{\mu\nu}\cr
	0&0&1\cr}
\end{equation}

\noindent acting for instance on the 11-dimensional colum vector $\pmatrix{ X^\mu&\cr\theta^{\mu\nu}&\cr 1&}$. The transformations (\ref{i19})
are reproduced for the infinitesimal case. In (\ref{P'}), $(\Lambda^{[\mu}_{\,\,\,\alpha}\Lambda^{\nu]}_\beta)$ forms  the antisymmetric product ($6\times6$) representation for $L$. With this structure we see that the usual
 classification scheme of the elementary particles accordingly to the eigenvalues of ${\mathbf C}_1$ and  ${\mathbf C}_2$ is not lost. However, $P$ and not $P$' can be the symmetry group when a particular theory is considered. This is the case of the theories adopting the quantum conditions (\ref{04}) or the Seiberg-Witten version of noncommutative gauge theories   \cite{Amorim5}. In this situation $P$' is dynamically contracted to $P$. 

An important point is that due to (\ref{02}) the operator ${\mathbf x}^\mu$ can not be used to label possible basis in ${\cal H}$. However, as the components of ${\mathbf X}^\mu$ commute, as can be verified from (\ref{i5}), their eigenvalues  can be used for such purpose. To simplify the notation, let us denote by $x$ and $\theta$ the eigenvalues of ${\mathbf X}$ and ${\mathbf\theta}$ in what follows. In   \cite{Amorim4} we have considered these points with some detail and have proposed a way for constructing some actions representing possible field theories in this extended $x+\theta$ space-time.  One of such actions has been given by

\begin{equation}
\label{b8}
S=-\int d^{4}\,x\,d^{6}\theta\,\Omega(\theta)\, {1\over2}\Big\{\,\partial^\mu\phi\partial_\mu\phi + \,{{\lambda^2}\over4}\,\partial^{\mu\nu}\phi\partial_{\mu\nu}\phi   +m^2\,\phi^2\Big\}\, 
\end{equation}

\noindent where $\lambda$ is a parameter with dimension of length, as the Planck length, which has to be introduced by dimensional reasons and $\Omega(\theta)$
is a scalar weight function used in   \cite{Carlson}-  \cite{Saxell} in order to make the connection between the $D=4+6$
and the $D=4$ formalisms. Also $\Box= \partial^\mu\partial_\mu  $, with $\partial_\mu={{\partial}\over{\partial {x}^\mu}}$ and  $\Box_{\theta}={1\over2}\partial^{\mu\nu}\partial_{\mu\nu}$,
where $\partial_{\mu\nu}={{\partial\,\,\,}\over{\partial {\theta}^{\mu\nu}}}\,\,$.  $\eta^{\mu\nu}=diag(-1,1,1,1)$.
\bigskip

The corresponding Lagrange equation reads

\begin{eqnarray}
\label{c6}
{{\delta S}\over{\delta\phi}}&=&\,\Omega\,(\Box - m^2)\phi+{{\lambda^2}\over2}\partial_{\mu\nu}(\Omega\,\partial^{\mu\nu}\phi)\nonumber\\
&=&\,\,\,0
\end{eqnarray}

\noindent and the action (\ref{b8}) is invariant under the transformation

\begin{equation}
\label{i19c}
\delta \phi=-(a^\mu+\omega^\mu_{\,\,\,\nu}x^\nu)\,\partial_\mu\phi-{1\over2}(b^{\mu\nu}+2\omega^\mu_{\,\,\,\rho}\theta^{\rho\nu})\,\partial_{\mu\nu}\phi
\end{equation}

\noindent when $\Omega$ is considered as a constant. If $\Omega$ is a non constant scalar function of $\theta$, the above transformation is only a symmetry of (\ref{b8}) when $b^{\mu\nu}$ vanishes, what dynamically contracts $P$' to $P$   \cite{Amorim4}. We observe that 
\noindent (\ref{i19c}) closes in an algebra, as in (\ref{xxx}), with the same composition rule defined in (\ref{i19b}). That equation defines  how a scalar field transforms in the $x+\theta$ space under the action of ${\cal P}'$. 

In what follows we are going to show how to introduce fermions in this $x+\theta$ extended space. To reach this goal, let us first observe that
$\cal P$' is a subgroup of the Poincar\'{e} group $\cal P$ $_{10}$ in $D=10$. Denoting the indices $A,B,...$ as space-time indices in $D=10$, $A,B,..=0,1,...,9$, a vector $Y^A$ would transform under $\cal P$ $_{10}$ as $\delta Y^A=\omega^A_{\,\,\,B}Y^B+ \Delta^A$, where the $45$ $\omega$'s and $10$ $\Delta$'s are infinitesimal parameters. If one identifies the last six $A,B,..$ indices with the macro-indices $\mu\nu$, $\mu,\nu,..=0,1,2,3$, considered as antisymmetric quantities, the transformation relations given above are rewritten as 

\begin{eqnarray}
\label{81}
\delta Y^\mu&=&\omega^\mu_{\,\,\,\nu}Y^\nu+{1\over2}\omega^\mu_{\,\,\,\alpha\beta}Y^{\alpha\beta}+\Delta^\mu\nonumber\\
 \delta Y^{\mu\nu}&=&\omega^{\mu\nu}_{\,\,\,\,\,\,\,\alpha}Y^\alpha+{1\over2}\omega^{\mu\nu}_{\,\,\,\,\,\alpha\beta}Y^{\alpha\beta}+\Delta^{\mu\nu}
\end{eqnarray}

With this notation, the ( diagonal ) $D=10$ Minkowski metric is rewritten as $\eta^{AB}=(\eta^{\mu\nu},\eta^{\alpha\beta,\gamma\delta})$ and the ordinary Clifford algebra $\{\Gamma^A,\Gamma^B\}=-2\eta^{AB}$  as

\begin{eqnarray}
\label{82}
\{\Gamma^\mu,\Gamma^{\alpha\beta}\}&=&0\nonumber\\
\{\Gamma^\mu,\Gamma^\nu\}&=&-2\eta^{\mu\nu}\nonumber\\
\{\Gamma^{\mu\nu},\Gamma^{\alpha\beta}\}&=&-2\eta^{\mu\nu,\alpha\beta} 
\end{eqnarray}

This is just a cumbersome way of writing  usual $D=10$ relations   \cite{Strings}. Now,
by identifying $Y^A$ with $(x^\mu,{1\over\lambda}\theta^{\alpha\beta})$, where $\lambda$ is some parameter with length dimension, we see from the structure given above  that the allowed transformations in   $\cal P$' are  those of  $\cal P$ $_{10}$, submitted to the conditions 

\begin{eqnarray}
\label{83}
\omega^{\mu\nu}_{\,\,\,\alpha}&=&\omega_{\mu\nu}^{\,\,\,\alpha}=0 \nonumber\\
\omega^{\mu\nu}_{\,\,\,\alpha\beta}&=&4\,\omega^{[\mu}_{\,\,\,\,\alpha}\delta^{\nu]}_{\,\,\,\beta}\nonumber\\
\Delta^\mu&=&a^\mu\nonumber\\
\Delta^{\alpha\beta}&=&{1\over\lambda}b^{\alpha\beta}\nonumber\\
\end{eqnarray}

\noindent obviously keeping the identification between $\omega^{AB}$ and $\omega^{\mu\nu}$ when $A=\mu$ and $B=\nu$. Of course we have now only $6$ independent $\omega$'s and $10$ $a$'s and $b$'s. With the relations given above it is possible to extract the "square root" of the generalized Klein-Gordon 
equation (\ref{c6}) 

\begin{equation}
\label{84}
(\Box+\lambda^2\Box_\theta-m^2)\phi=0
\end{equation}

\noindent assuming here that $\Omega$ is a constant. This gives just the generalized Dirac equation

\begin{equation}
\label{85}
\Big[ i(\Gamma^\mu\partial_\mu+{\lambda\over2}\Gamma^{\alpha\beta}\partial_{\alpha\beta})-m\Big]\psi=0
\end{equation}

Actually, by applying  from the left by the operator 
$\Big[ i(\Gamma^\nu\partial_\mu+{\lambda\over2}\Gamma^{\alpha\beta}\partial_{\alpha\beta})+m\Big]$ on (\ref{85}),
 after using (\ref{82}) we observe that $\psi$ satisfies the generalized Klein-Gordon equation (\ref{84}) as well. The covariance of the generalized Dirac equation (\ref{85}) can also be proved. Fist we note that the operator

\begin{equation}
\label{86}
M^{\mu\nu}={i\over4}\Big([\Gamma^\mu,\Gamma^\nu]+[\Gamma^{\mu\alpha},\Gamma^\nu_{\,\,\,\alpha}]\Big)
\end{equation}

\noindent gives the desired representation for the $SO(1,3)$ generators, because it not only closes in the Lorentz algebra (\ref{i17}), but also satisfies the commutation relations

\begin{eqnarray}
\label{87}
&&[\Gamma^\mu,M_{\alpha\beta}]=2i\delta^{\mu}_{[\alpha}\Gamma_{\beta]}\nonumber\\
&&[\Gamma^{\mu\nu},M_{\alpha\beta}]=2i\delta^{\mu}_{[\alpha}\Gamma^{\,\,\,\nu}_{\beta]}-2i\delta^{\nu}_{[\alpha}\Gamma^{\,\,\,\mu}_{\beta]}
\end{eqnarray}

With these relations it is possible to prove that (\ref{85}) is indeed covariant under the Lorentz transformations given by 

\begin{equation}\label{88}
\psi(x',\theta')=exp(-{i\over2}\Lambda^{\mu\nu}M_{\mu\nu})\psi(x,\theta)
\end{equation}

By considering the complete $\cal P$' group, we observe that the infinitesimal transformations of $\psi$ are given by

\begin{equation}\label{89}
\delta\psi=-\Big[(a^\mu+\omega^\mu_{\,\,\,\nu}x^\nu)\partial_\mu+{1\over2}(b^{\mu\nu}+2\omega^\mu_{\,\,\,\rho}\theta^{\nu\rho})\partial_{\mu\nu}+{i\over2}\omega^{\mu\nu}M_{\mu\nu}\Big]\psi
\end{equation}

\noindent which closes in the $\cal P$' algebra with the same composition rule given by (\ref{i19b}), what can be shown after a little algebra. At last we can show that also here there are conserved Noether's currents associated with the transformation (\ref{89}), once we observe that the equation (\ref{85})
can be derived from the action

\begin{equation}
\label{90}
S=\int d^{4}\,x\,d^{6}\theta\,\Omega(\theta)\, \bar \psi \Big[ i(\Gamma^\mu\partial_\mu+{\lambda\over2}\Gamma^{\alpha\beta}\partial_{\alpha\beta})-m\Big]\psi
\end{equation}

\noindent where  we are considering $\Omega=\theta_0^{-6}$ and $\bar \psi=\psi^\dag\Gamma^0$. First  we note that ( suppressing trivial $\theta_0^{-6}$ trivial factors )

\begin{eqnarray}
\label{91}
{{\delta^L S}\over{\delta\bar\psi}}&=&\Big[ i(\Gamma^\mu\partial_\mu+{\lambda\over2}\Gamma^{\alpha\beta}\partial_{\alpha\beta})-m\Big]\psi\nonumber\\
{{\delta^R S}\over{\delta\psi}}&=&-\bar\psi\Big[ i(\Gamma^\mu\overleftarrow\partial_\mu+{\lambda\over2}\Gamma^{\alpha\beta}\overleftarrow\partial_{\alpha\beta})+m\Big]
\end{eqnarray}

\noindent where $L(R)$ derivatives act from the left(right). The current $(j^\mu,j^{\mu\nu})$, as in   \cite{Amorim4}, is here written as

\begin{eqnarray}
\label{92}
&&j^\mu={{\partial^R {\cal L}}\over{\partial\partial_\mu\psi}}\delta\psi+\delta\bar\psi {{\partial^L {\cal L}}\over{\partial\partial_\mu\bar\psi}}+
{\cal L}\delta x^\mu\nonumber\\
&&j^{\mu\nu}={{\partial^R {\cal L}}\over{\partial\partial_{\mu\nu}\psi}}\delta\psi+
\delta\bar\psi {{\partial^L{\cal L}}\over{\partial\partial_{\mu\nu}\bar\psi}}+{\cal L}\delta \theta^{\mu\nu}
\end{eqnarray}

\noindent where

\begin{equation}\label{93}
\delta\bar\psi=-\bar\psi\Big[\overleftarrow\partial_\mu(a^\mu+\omega^\mu_{\,\,\,\nu}x^\nu)+\overleftarrow\partial_{\mu\nu}{1\over2}(b^{\mu\nu}+2\omega^\mu_{\,\,\,\rho}\theta^{\nu\rho})-{i\over2}\omega^{\mu\nu}M_{\mu\nu}\Big]
\end{equation}

\noindent $\delta\psi$ is given by (\ref{89}) and $\delta x^\mu$ and $\delta\theta^{\mu\nu}$ have the same form found in (\ref{i19}). After a long but direct calculation one can show that

\begin{equation}\label{94}
\partial_\mu j^\mu+\partial_{\mu\nu}j^{\mu\nu}=-\Big(\delta\bar\psi{{\delta^L S}\over{\delta\bar\psi}}+{{\delta^R S}\over{\delta\psi}}\delta\psi\Big)
\end{equation}

\noindent which vanishes on shell, proving the invariance of the action (\ref{90}) under $\cal P$'. By the reasons pointed through this work, it could be dynamically contracted to $P$, preserving the usual Casimir invariant structure characteristic of ordinary quantum field theories.

Due to (\ref{94}) there is a conserved charge

\begin{equation}
\label{95}
Q=\int d^3 x d^6 \theta\, j^0
\end{equation}

\noindent for each one of the specific transformations encoded in (\ref{92}). Actually, $\dot Q=-\int d^3 x d^6 \theta\, (\partial_i\,j^i+\partial_{\mu\nu}j^{\mu\nu})$
vanishes as a consequence of the divergence theorem. By considering only $x^\mu$ translations, we can write $j^0=j^0_\mu a^\mu$, permitting to define the
momentum operator $P_\mu=-\int d^3x d^6\theta j^0_\mu$. Also by considering $\theta^{\mu\nu}$ translations and Lorentz transformations, we can derive in a similar way an explicit form for the other generators of $\cal P$', here denoted by $\Pi_{\mu\nu}$ and $J_{\mu\nu}$. Under an appropriate bracket structure, as teaches us the Noether's theorem, these conserved charges will generate the transformations (\ref{89}) and (\ref{93}).

We close this work by observing that we have been able to introduce fermions satisfying a generalized Dirac equation, which is covariant under the action of the extended Poincar\'{e} group $P$'. That equation has been derived through a variational principle whose action is dynamically invariant under $P$'. This can clarify possible rules played by theories involving noncommutativity in a way compatible with Relativity. Of course this is just a little step toward a field theory quantization program in this extended $x+\theta$ space-time. This last point is under study and possible results will be reported elsewhere.


\begin{thebibliography}{30}
\bibitem{NCQM} In Refs.   \cite{Amorim1} and     \cite{Amorim4} it is possible to find a large amount of references concerning noncommutative quantum mechanics.
\end{thebibliography}
\end{document}